\documentclass[11pt]{article}
\usepackage{fullpage}
\usepackage{amsfonts}
\usepackage{amsmath, amsthm, amssymb}
\usepackage[inline]{enumitem}
\usepackage{mathtools}
\usepackage{multicol}
\usepackage{booktabs,tabularx}
\usepackage{setspace}
\usepackage{subcaption}
\usepackage{float}
\usepackage{setspace}
\usepackage{color}
\usepackage{soul}
\usepackage{xfrac}
\usepackage{hyperref}
\usepackage{cleveref}
\usepackage{lineno}
\usepackage{placeins}
\usepackage{lmodern}
\usepackage{authblk}

\title{Altruism and energy flow in dynamic beehive models}

\date{\today}
\author{Zachary Nathan\thanks{Department of Computer Science, Dartmouth College} , Daniel DiPietro\thanks{Departments of Mathematics and Computer Science, Dartmouth College} , and Olivia J. Chu\thanks{Department of Mathematics \& Neukom Institute for Computational Science, Dartmouth College and Department of Mathematics, Bryn Mawr College} \thanks{ochu@brynmawr.edu}}

\begin{document}

\maketitle

\begin{abstract}

    This work explores the relationship between altruism and the genetic system of arrhenotoky through an evolutionary game theory (EGT)-inspired lens, using a dynamic model of beehive populations consisting of three castes: workers, drones, and the queen. Arrhenotoky is a form of asexual reproduction in which unfertilized eggs become males while fertilized eggs develop into females, leading to unusual patterns of genetic relatedness between family members. This mode of reproduction occurs in insects such as the \textit{Hymenoptera}, including bees. In the hive environment, bees often display altruistic behavior, or actions taken by an organism that reduce its own fitness to increase the fitness of others. Eusociality, an elaborate form of social organization characterized by complex and altruistic social behaviors, is also observed in the \textit{Hymenoptera}. To explore the interplay between altruism and the reproductive patterns of arrhenotoky, we employ a population dynamics model to simulate beehive populations over a range of parameters, controlling for altruism in workers and the queen. Our results show that altruistic behaviors are essential for beehive success, with optimal worker altruism corresponding to the division of labor observed in eusocial species. Furthermore, we find that modest altruism from the queen is also vital for hive survival, emphasizing the delicate balance that can exist in these complex social systems. Overall, our findings shed light on the co-evolution of altruism, arrhenotoky, and eusociality in the natural world.
    
\end{abstract}

\newpage

\section{Introduction}

% Comment about how bees are interesting 

Social insects -- bees, wasps, termites, ants -- have long puzzled researchers with their altruistic behavior. Within these insect communities, life is exceptionally organized, as each member knows and carries out his or her respective tasks. Most individuals perform non-reproductive duties in support of the one or few reproducing members. This begs the question of how this arrangement is compatible with natural selection and ``survival of the fittest," and in particular, the role that genetics plays (or does not play) in the altruistic behaviors observed in social insect colonies. 
 
In this work, we investigate the connection between altruism and arrhenotoky, the form of reproduction that occurs in several social insect populations, using a population dynamics model based on the flow of energy. We develop a dynamic population model in which individuals are divided into castes based on their roles. Individuals in each of these groups allocate energy within and between them according to variable levels of altruism. The energy surpluses within each caste then determine their respective capacities to reproduce. Reproduction occurs in a three-part balance according to the biological system of arrhenotoky (\Cref{fig:arrhenotoky}).

Although our model can describe any species exhibiting arrhenotoky, we focus our analysis on the honey bee, \textit{Apis Mellifera}. Honey bees naturally exhibit arrhenotoky along with complex social structures, making them prime candidates for analyzing the effects of altruism \cite{goudie_2018}. Honey bees are essential to ecosystems and agriculture worldwide; in the year 2000, \textit{Apis Mellifera} contributed an estimated \$14.6 billion via pollination in the United States alone \cite{morse_2000}. Furthermore, a wealth of research has been published on the genomics of bees, giving insight into the genetic basis of arrhenotoky and altruism \cite{mcafee_2019,leonard_2020,grozinger_2020}.

%Finally, we note that bees are difficult to study at the individual level in lab settings, when compared to flightless species with arrhenotoky such as ants \cite{kimura_2014}, making a mathematical analysis of their behavior particularly useful.

\subsection{Beehive Dynamics} \label{biobackground}

A beehive typically houses between $10,000$ and $100,000$ individuals, depending on the species and environment. In all cases, these populations are split into three distinct castes: a female queen, female workers, and male drones \cite{rowland_1987}.

The queen bee is the leader of the hive, and the mother of most if not all of the other bees. With few exceptions, beehives have a single adult queen who is its largest and longest-living member. She is always surrounded by worker bees, who provide her with food as well as physical protection. The queen's primary purpose is to reproduce; with access to enough food and mates, she can lay thousands of eggs per day \cite{ratnieks_2009}. These eggs hatch into workers if fertilized, or drones if not.

The main role of drone bees is to mate with the queen, enabling the birth of workers via fertilization. As soon as a drone mates with a queen, it dies. Drones are born from unfertilized eggs laid by the queen or workers. Drones constitute anywhere from one to twenty percent of the total beehive population, although this varies seasonally and between species. Typically, they mate with queens from other hives to promote genetic diversity, gathering in drone congregation areas \cite{ayup_2021}. The secondary role of drones is temperature regulation; their populations peak in the summer, but during the winter when the hive's resources are limited, they die off or get evicted \cite{rowland_1987}. 

Worker bees are responsible for sustaining the hive, performing vital functions such as: gathering energy by collecting nectar and pollen (food), nursing larvae, producing wax and honeycomb, attending to the queen, and defending the hive from invaders. Workers can lay eggs of their own, but these are never fertilized; therefore, workers alone are only capable of producing more drones. This is rare, however, as the vast majority of honeybee drones are the queen's direct offspring \cite{ratnieks_2009}. Workers themselves are born from eggs that are laid by the queen and fertilized by drones.

New queen bees are born in a process similar to that of the workers, hatching from fertilized eggs. However, queen larvae are exclusively fed `royal jelly', a protein-rich secretion that enables their development to sexual maturity. In contrast, worker larvae are only fed royal jelly for their first few days, after which they're fed a combination of nectar and pollen known as `bee bread'. The queen larvae selection mechanism is not yet fully understood \cite{zhu_2017}. Once the new queen bee reaches maturity, one of two processes typically occurs. If the old queen no longer releases sufficient pheromones, she is killed by the workers, making way for the new queen in a process called supersedure. Otherwise, the hive undergoes a split; through a process known as swarming, the old queen founds a new hive, taking roughly half of the population along with her \cite{rowland_1987}.

\subsection{Arrhenotoky}\label{arrhenotoky}

\begin{figure}%[hbt!]
     \centering
     \includegraphics[width=0.7\linewidth]{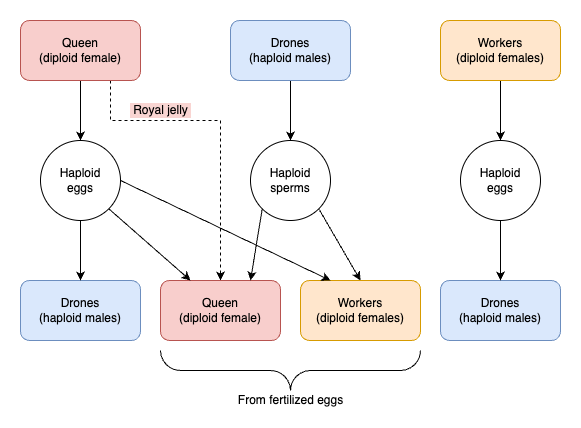}
     \caption{A visual overview of honey bee reproduction via arrhenotoky. Fertilized eggs become diploid females, whereas unfertilized eggs become haploid males. Queens are distinguished from workers by the provision of royal jelly during the larval stage \cite{zhu_2017}.}
     \label{fig:arrhenotoky}
 \end{figure}

Arrhenotoky, or arrhenotokous parthenogenesis, is a form of asexual reproduction distinguished by the haplodiploid sex-determination system. Specifically, unfertilized eggs develop into haploid male offspring, while fertilized eggs develop into diploid female offspring having twice the number of chromosomes (\Cref{fig:arrhenotoky}). This phenomenon is characteristic of the \textit{Hymenoptera}, an order of insects including bees, ants, wasps, and sawflies, as well as the \textit{Thysanoptera}, or thrips. Species with arrhenotoky often exhibit distinct genetic and social dynamics, with complex group behaviors including eusociality \cite{goudie_2018}.

With arrhenotoky, the distinctive pathways of reproduction lead to higher degrees of relatedness among female siblings, than between parents and offspring. In particular, diploid workers share more genetic material with one another ($\sim$75\%) than with their own haploid drone offspring (50\%) \cite{ratnieks_2009}. The queen is the mother of all workers as well as most drones; hence, workers inherit half of their genes directly from the queen, and the other half from drones who (almost always) inherit from the queen in turn. Therefore, workers in a hive share 75\% of their genes, or marginally less in the rare case that a drone father was hatched from another worker's egg. This peculiar genetic arrangement gives workers an incentive to act altruistically, as their genes are propagated more effectively via the queen's reproduction than their own \cite{ratnieks_2009}. Furthermore, the queen may act altruistically by supporting the drones, who need to survive until fertilization to enable the reproduction of workers. Hence, arrhenotoky gives rise to altruism in two distinct forms: workers helping the queen, and the queen helping drones.

We emphasize that in our model, altruism is defined in terms of energy allocation. Within the hive, energy flows in a natural way: from workers (who gather energy on behalf of the hive) to the queen (who is surrounded by workers) to drones (who do not lay eggs and who wait to mate with a queen before dying). This set-up does not consider the relatedness coefficients between members of the hive.

\subsection{Altruism}

In an evolutionary context, altruism is defined as any action taken by an organism that reduces its own fitness in order to increase that of others. Cooperative breeding or hunting, the sharing of food or shelter, and predator alarm calls are all examples of altruism in the animal kingdom \cite{altruism}. Understanding the mechanisms that drive altruistic behavior has significant implications for the study of social organisms and their interactions. Various theories may describe the evolution of altruism, as discussed in \Cref{rw}.

Perhaps the most extreme form of biological altruism is eusociality, which describes the advanced social organization observed in the \textit{Hymenoptera} and several other species, with and without arrhenotoky \cite{strassmann_2011}. Eusocial behavior is defined by the division of labor into reproductive and non-reproductive castes, cooperative care of offspring, and overlapping generations of adults \cite{hughes_2008}. Such advanced cooperation allows for colonies to operate much more effectively than the individuals' intelligence and abilities would otherwise permit. The \textit{Hymenoptera}, including bees, exhibit both arrhenotoky and eusociality, rendering them compelling objects of study \cite{ratnieks_2009}.

\section{Related Work}\label{rw}

The evolution of altruism and, in particular, the eusocial behaviors exhibited by certain insects, represent a fascinating question in the field of evolutionary biology. A significant body of research supports the theory of inclusive fitness, or kin selection, as the primary driving force behind these behaviors. Hamilton, in his seminal work, proposed the concept of kin selection, suggesting that organisms may increase their own genetic success by aiding relatives, forming the basis of altruistic behavior \cite{hamilton_1963}. Looking at insects, Hughes et al. and Boomsma have presented findings suggesting that monogamy serves as the foundation for the evolution of eusociality, by maximizing relatedness among descendants \cite{hughes_2008,boomsma_2009}. Further studies by Strassman et al., Liao et al., and Queller et al. have reinforced the validity of kin selection as an explanation for eusociality, highlighting the significance of intragenomic conflict \cite{strassmann_2011,liao_2015,queller_2015}. Galbraith et al. provided empirical support for this theory, demonstrating how kin selection predicts intragenomic conflict in honey bees \cite{galbraith_2016}.

Alternative theories challenging the primacy of kin selection in explaining the evolution of eusociality have also emerged. For instance, Ratnieks and Helantera argued that coercion better explains the extreme altruism and social inequalities observed in insect societies \cite{ratnieks_2009}. A paradigm shift was introduced by Nowak et al., who proposed a five-step sequence to explain the evolution of eusociality. Their theory suggested that eusociality arises through \begin{enumerate*}[label=(\roman*)]\item group formation, \item accumulation of specialized traits, \item evolution of eusocial alleles, and \item natural selection of emergent traits, leading to \item between-colony selection shaping life cycles and caste systems \end{enumerate*} \cite{nowak_2010}. Nowak and Allen later contended that the phenomena invoked in inclusive fitness theorizing were not relevant to eusociality \cite{nowak_2015}.

This ongoing debate highlights the confounding nature of altruism in evolutionary biology, especially in the case of eusocial insects. The inclusive fitness/kin selection theory as well as the alternatives contribute valuable insights toward understanding the evolution of these phenomena, while encouraging further study. Our model does not presuppose any theories in particular, but constitutes an entirely separate analytical pathway.

Evolutionary game theory (EGT) and population dynamics models provide a useful interdisciplinary framework for the analysis of varied dynamical systems \cite{traulsen_2023}. Our work takes inspiration from several  studies that have modeled biological systems using  simulations and analysis \cite{hammerstein_2015}. These studies can model the dynamics of cooperation and competition, even among organisms below the level of cognition \cite{hummert_2014}, from altruism in honeybees to facultative cheating in yeast \cite{gore_2009}. Other similar studies focus on the dynamics of population growth \cite{melbinger_2010}, as well as the stability \cite{cavaliere_2012} and structure \cite{ohtsuki_2006} of networks of populations.

\section{Methods}

We present an ordinary differential equation (ODE) model of arrhenotoky in the beehive setting. The model simulates beehive populations over a grid of worker/queen altruism configurations.

\subsection{Model}\label{model}

\begin{figure}%[H]
    \centering
    \includegraphics[width=1.0\linewidth]{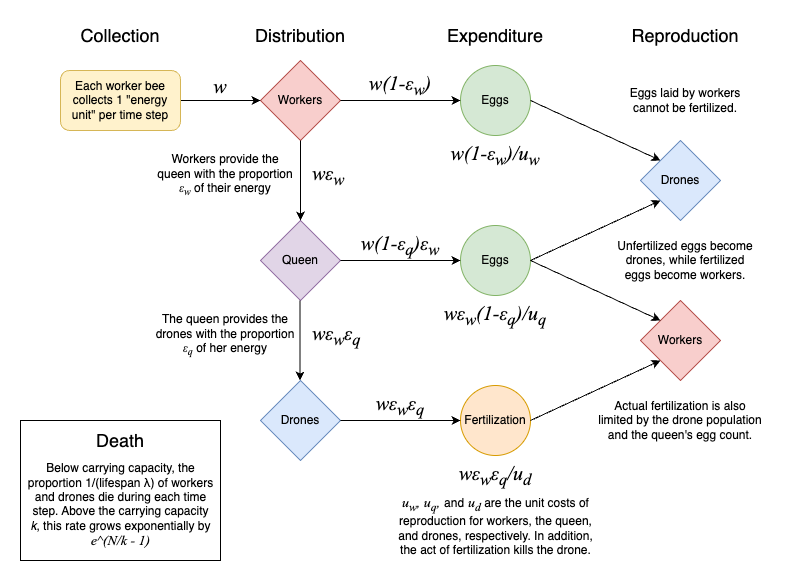}
    \caption{Model overview, illustrating the flow of energy from collection through reproduction. $N$ = total population size; $w$ = number of workers; $\varepsilon$ = altruism coefficient; $u$ = unit cost of reproduction; $k$ = carrying capacity; $\lambda$ = lifespan. Altruism occurs in the distribution stage. Arrow labels denote energy amounts, while expenditure labels indicate capacity for reproduction.}
    \label{fig:model}
\end{figure}

Here, we introduce our mathematical model of arrhenotoky reproductive dynamics. This model is formulated within a population dynamical framework as illustrated in \Cref{fig:model}. The workers gather energy, keeping a portion for themselves and altruistically donating the rest to the queen. The queen, in turn, keeps a portion of this energy for herself, donating the rest to the drones. We offer further mathematical details below.

\subsubsection{Energy Payoffs}

Suppose that each worker bee gathers $1$ net unit of energy for the hive at each time step, on average. They provide a proportion $\varepsilon_w \in [0,1]$ of this resource to the queen and keep the remaining $(1-\varepsilon_w)$ for themselves. Therefore, the worker population has the following energy payoff per time step, where $w$ is the number of workers:
\begin{equation}\label{eq:workerenergy}
    J_w = w (1-\varepsilon_w)
\end{equation}
Now, the remaining energy $w\varepsilon_w$ is provided by the workers to the queen. Much like the workers, she keeps a proportion $(1-\varepsilon_q)$ of this energy for herself, releasing the remaining $\varepsilon_q \in [0,1]$ to be shared amongst the drones. So, the queen has the following energy payoff each round:
\begin{equation}\label{eq:queenenergy}
    J_q = w (1-\varepsilon_q) \varepsilon_w 
\end{equation}
Finally, the collective drone population is left with the following payoff:
\begin{equation}\label{eq:droneenergy}
    J_d = w \varepsilon_w \varepsilon_q
\end{equation}

\subsubsection{Reproduction}

As detailed in \Cref{biobackground}, arrhenotoky dictates that each caste plays a distinct role in the reproductive process:
\begin{itemize}
    \item Workers produce eggs that develop into drones
    \item The queen produces eggs that develop into workers if fertilized, or drones if not
    \item Drones mate with the queen, fertilizing her eggs and dying in the process
\end{itemize}

% Comment about how we model DCAs 
% Our model, however, considers beehive population dynamics in general, abstracting away the mating locations and genetics. We can equivalently think of this as modeling the extended neighborhood of a central hive.

In order to account for drones mating with queens from different hives in drone congregation areas, as described above, we consider an abstract  `extended neighborhood' of a focal hive, that includes drones not genetically related to the queen. 

We denote the unit costs of reproduction, in terms of energy, as $u_w$, $u_q$, and $u_d$ for the three respective castes. While the real-world energy requirements may differ between species, we assert that $u_d<u_q<u_w$. Given that queen bees lay more eggs than workers by several orders of magnitude \cite{ratnieks_2009}, it follows that their process would be more efficient in terms of the energy expenditure per unit. The low relative cost of drone reproduction arises from Bateman's principle \cite{bateman_1948}; notwithstanding any flaws in his original experiment, we recognize that `sperm is cheap' in comparison to eggs \cite{olsson_1997}. Importantly, this unit cost does not take the drone's ensuing death into account.

Using these unit costs of reproduction along with the energy payoffs for each caste, we proceed to calculating their respective capacities for reproduction. First, we define $E_q$ as the number of unfertilized eggs produced by the queen:
\begin{equation}\label{eq:queenrep}
    E_q = \frac{J_q}{u_q}
\end{equation}
Similarly, we define $E_w$ as the number of eggs collectively produced by the workers:
\begin{equation}\label{eq:workerrep}
    E_w = \frac{J_w}{u_w}
\end{equation}
Finally, we define $E_d$ as the number of eggs collectively fertilized by the drones. This is slightly more complicated, as drones cannot fertilize more eggs than have been produced by the queen, or more eggs than the current number of drones, as each one dies after fertilization. So, where $d$ is the number of drones, we define $E_d$ as:
\begin{equation}\label{eq:dronerep}
    E_d = \min\left(E_q, d, \frac{J_d}{u_d} \right)    
\end{equation}

\subsubsection{Death}\label{death}

Whereas queen bees are relatively long-lived, workers and drones die with regularity throughout the lifespan of the hive \cite{ratnieks_2009}. Thus, we extend our model to include death dynamics, using the hive's carrying capacity along with expected lifespans for workers and drones. Suppose that workers and drones die after $\lambda_w$ and $\lambda_d$ time steps, on average, respectively. In order to achieve these average lifespans, a proportion $\frac{1}{\lambda}$ of each respective caste dies at each time step. That is, $\frac{w}{\lambda_w}$ workers and $\frac{d}{\lambda_d}$ drones die of `natural causes', as long as the total population size $N$ remains below the carrying capacity $k$.

The carrying capacity $k$ represents the number of bees $N$ that can live safely within a single hive, before excess deaths begin to occur due to resource competition, overcrowding, predation, or other constraints \cite{al-ghamdi_2016}. As long as $N\leq k$, the rate of natural deaths remains unchanged. To enforce the carrying capacity, we increase the rate of natural deaths (when $N > k$) by the exponential factor $e^{\frac{N}{k}-1}$, determined by the degree to which $N$ exceeds $k$, allowing for realistic population totals above the carrying capacity.

We define $D_w$ as the number of worker deaths:
\begin{equation}\label{eq:workerdeath}
    D_w = \max\left( 1, e^{\frac{N}{k}-1} \right) \frac{w}{\lambda_w}
\end{equation}
Similarly, we define $D_d$ as the number of drone deaths, recalling that $E_d$ drones die as a result of fertilization:
\begin{equation}\label{eq:dronedeath}
    D_d = E_d + \max\left( 1, e^{\frac{N}{k}-1} \right) \frac{d}{\lambda_d}
\end{equation}

\subsubsection{Replicator Dynamics}

Worker and drone demographic equations are written as ODEs, in the general form:
\begin{equation}
    \dot{P} = \frac{dP}{dt} = \texttt{births} - \texttt{deaths}
\end{equation}
for a population $P$.
 
We define these demographic equations below, separately for workers and then drones, based on their respective birth and death rates as previously discussed. First, we define $\dot{w}$ for the workers, recalling that they're born from fertilized eggs ($E_d$):
\begin{equation}\label{wdot}
    \begin{split}
        \dot{w} &= E_d - D_w\\
        \dot{w} &= \min\left(\frac{w (1-\varepsilon_q) \varepsilon_w}{u_q}, d, \frac{w \varepsilon_w \varepsilon_q}{u_d} \right) - \max\left( 1, e^{\frac{N}{k}-1} \right) \frac{w}{\lambda_w}
    \end{split}
\end{equation}
Next, we define $\dot{d}$ for the drones, recalling that they're born from unfertilized queen eggs ($E_q - E_d$) as well as worker eggs ($E_w$):
\begin{equation}\label{ddot}
    \begin{split}
        \dot{d} &= E_w + (E_q - E_d) - D_d\\
        \dot{d} &= \frac{w (1-\varepsilon_w) }{u_w} + \frac{w (1-\varepsilon_q) \varepsilon_w }{u_q} - 2\min\left(\frac{w (1-\varepsilon_q) \varepsilon_w}{u_q}, d, \frac{w \varepsilon_w \varepsilon_q}{u_d} \right) - \max\left( 1, e^{\frac{N}{k}-1} \right) \frac{d}{\lambda_d}
    \end{split}
\end{equation}

\subsection{Simulations}

To observe the effects of altruism, we simulate the beehive model in a grid search over the independent variables $\varepsilon_w$ and $\varepsilon_q$, which represent altruism coefficients from 0 (total selfishness) to 1 (total altruism) for the workers and queen respectively. The simulation constants are defined in \Cref{tab:constants} and estimated from the literature when possible. Generally, these constants do not affect the model's behavior; adjusting their values serves only to shift boundaries and scale results, as shown in the Supplementary Information (\Cref{supplement}). For example, doubling $k$ will simply double the equilibrium populations for all $\varepsilon_w$ and $\varepsilon_q$. As such, it makes sense to discuss the resulting populations in terms of $k$.

\begin{table}[h]
    \centering
    \caption{Simulation constants}
    \label{tab:constants}
    \begin{tabular}{l c l}
        \toprule
        Symbol & Value & Description \\
        \midrule \midrule
        $k$ & 10000 & Carrying capacity \\
        $T$ & 10000 & Simulation time span \\
        $w_0$ & 30 & Initial worker population \\
        $d_0$ & 5 & Initial drone population \\
        $\lambda_w$ & 45 & Average worker lifespan \\
        $\lambda_d$ & 45 & Average drone lifespan \\
        $u_w$ & 3.0 & Worker reproduction unit cost \\
        $u_q$ & 1.5 & Queen reproduction unit cost \\
        $u_d$ & 0.5 & Drone reproduction unit cost \\
        \bottomrule
    \end{tabular}
\end{table}

\section{Results}

\subsection{Simulation Outcomes}

We present the simulation results in \Cref{fig:results}, illustrating the effects of altruism under arrhenotoky. The heatmaps cover the full range of $\varepsilon_w$ and $\varepsilon_q$ from $0$ to $1$, at increments of $0.0025$. We present three important results from each simulation of one pair $(\varepsilon_q,\varepsilon_w)$: the population total $N_T$, composition ratio $\sfrac{d}{w}$, and time to convergence $t_c$.

\subsubsection{Population Totals}

The population total $N_T$ refers to the number of workers $w$ plus the number of drones $d$ at time $t=T$, the end of the simulation. This total does not include the singular queen, whose static population is not affected by the ODEs. In \Cref{fig:results_totals}, we present these population totals for all 160,681 simulated pairs of $\varepsilon_q$ and $\varepsilon_w$.

The highest observed population is $34,479$ (or $3.45k$), occurring when $\varepsilon_q=0.13$ and $\varepsilon_w=1.0$. Similar populations exist along a line in the parameter space, in the southeast (negative $\varepsilon_w$ and positive $\varepsilon_q$) direction. We note the appearance of three high-population `ridges' in the shape of a wishbone, effectively dividing the parameter space into three regions. We explain these regions and derive their boundaries in \Cref{math}. Overall, high populations result from plausible levels of altruism across a wide range. Specifically, we observe $N_T>2k$ for most $0.1<\varepsilon_q<0.75$ and $\varepsilon_w>0.25$.

Interestingly, non-zero populations do not exist below the carrying capacity $k$; simulations with too low $\varepsilon_w$ or extreme $\varepsilon_q$ always converge to zero. We refer to these populations as non-viable, and the edge between them and their viable counterparts as the viability boundary. In our model, reproduction either \begin{enumerate*}[label=(\roman*)]\item outpaces death by natural causes until some equilibrium $N>k$, or \item fails before reaching the $k$ threshold, eventually falling to zero \end{enumerate*}. We explain the connection between $k$ and deaths in \Cref{death} and derive the viability boundary in \Cref{viability}.

\subsubsection{Population Compositions}

Given that beehive populations consist mainly of two distinct castes, workers and drones, we report population compositions $\sfrac{d}{w}$ in \Cref{fig:results_compositions}, alongside the totals for each simulation. Here, compositions are measured by the numbers of drones per worker at $t=T$. We plot \Cref{fig:results_compositions} with a logarithmic scale, in order to better capture the huge variations in $\sfrac{d}{w}$ across the parameter space. The observed population compositions $\sfrac{d}{w}$ range from $0.0219$ drones per worker (at $\varepsilon_q=0.9425$, $\varepsilon_w=0.98$) up to $26.5$ (at $\varepsilon_q=0.0125$, $\varepsilon_w=0.8625$).

A `wedge' shape is apparent in the upper half of this heatmap, with its edges matching the `ridges' observed in \Cref{fig:results_totals}. This suggests that $\sfrac{d}{w}$ is determined by the same underlying factors as $N_T$. Inside the wedge, population compositions exhibit relatively little variation, all having more workers than drones (that is, $\log\sfrac{d}{w}<0$). Moving outside of the wedge, drones gradually become more prevalent until reaching the viability boundary, beyond which the compositions $\sfrac{d}{w}$ are undefined. One exception is the upper-right corner (where $\varepsilon_q\approx0.9$ and $\varepsilon_w >0.9$) which has the fewest drones. Under such high levels of altruism, the scarce eggs are nearly all fertilized by overfed drones, yielding fewer drones in each subsequent generation.

For context, we recall that in real-world beehives, drones typically make up between one and twenty percent of the population, varying seasonally and between species  \cite{rowland_1987}. This corresponds to compositions $\sfrac{d}{w}$ between $0.01$ and $0.25$ (or $\log\sfrac{d}{w}$ between $-4.6$ and $-1.4$ as in \Cref{fig:results_compositions}). The entire wedge area falls within this range, as opposed to other areas having lower populations $N_T$.

In general, our model is flexible and can be adapted to describe, simulate, and analyze varied behaviors within a colony. One peculiar example is the rare `anarchistic' phenotype that has been observed in honeybee populations \cite{workersterility}. When this phenotype is present, workers, who seldom lay eggs, develop functional ovaries and lay many, which, unfertilized, become drones. While the norm is for workers to be mainly non-reproductive (corresponding to our `optimal' worker altruism near $\varepsilon_w = 1$), our model could easily be adapted for the presence of this phenotype. The resulting population dynamics can be analyzed by considering $\varepsilon_w < 1$, or even by altering the replicator dynamics that influence workers' reproductive capacities. 

\subsubsection{Times to Convergence}

\begin{figure}
    \centering
    
    \begin{subfigure}{.5\textwidth}
        \centering
        \captionsetup{width=.9\linewidth}
        \includegraphics[scale=0.5]{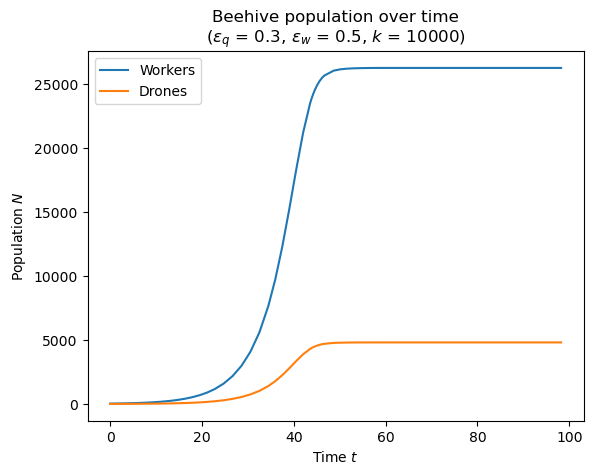}
        \caption{Population growth over time for a single hive with $\varepsilon_q=0.3$ and $\varepsilon_w=0.5$. Truncated at $t=100$ to highlight the growth curve.\\}
        \label{fig:results_single_simulation}
    \end{subfigure}%
    \begin{subfigure}{.5\textwidth}
        \centering
        \captionsetup{width=.9\linewidth}
        \includegraphics[scale=0.5]{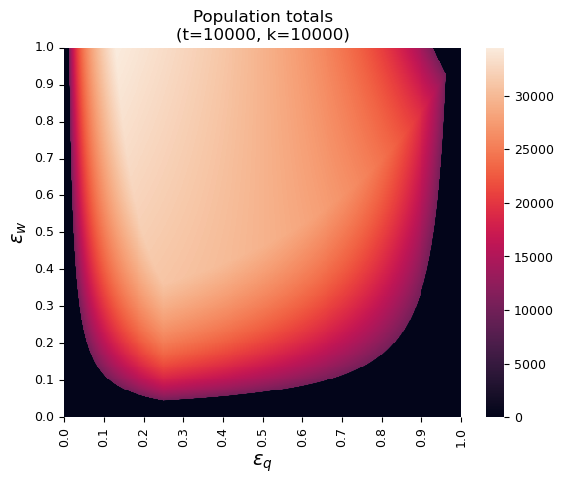}
        \caption{Heatmap showing the final population totals $N_T$ for all values of $\varepsilon_q$ (queen altruism) and $\varepsilon_w$ (worker altruism).\\}
        \label{fig:results_totals}
    \end{subfigure}

    \begin{subfigure}{.5\textwidth}
        \centering
        \captionsetup{width=.9\linewidth}
        \includegraphics[scale=0.5]{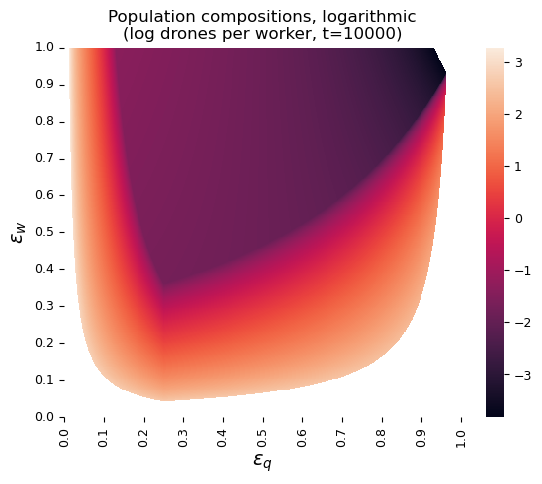}
        \caption{Heatmap showing population compositions (numbers of drones per worker) for non-zero populations. Logarithmic, to improve contrast and highlight the `wedge' area where workers outnumber drones.}
        \label{fig:results_compositions}
    \end{subfigure}%
    \begin{subfigure}{.5\textwidth}
        \centering
        \captionsetup{width=.9\linewidth}
        \includegraphics[scale=0.5]{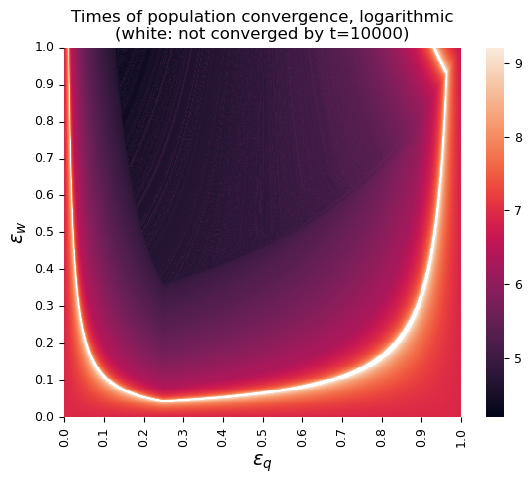}
        \caption{Heatmap showing times $t$ of population convergence at a steady state. Logarithmic to improve contrast. Note how populations along the viability boundary take the longest, or fail to converge by $t=10000$.}
        \label{fig:results_convergences}
    \end{subfigure}
    
    \caption{Simulation results for the full ranges of $\varepsilon_q$ and $\varepsilon_w$.}
    \label{fig:results}
\end{figure}

For certain values of $\varepsilon_q$ and $\varepsilon_w$, the simulated populations take vastly longer to converge to a steady state. In \Cref{fig:results_convergences}, we present the times to convergence $t_c$ for each simulation. The concept of convergence is intuitive, but an exact definition is tricky to formulate under the constraints of our model's implementation. Here, we define the time to convergence as the first time $t$ where $|N_t-N_{t-dt}|<10^{-5}$. The exact time increments $dt$ between consecutive simulation steps are varied, but guaranteed not to exceed $2.5$.

As in \Cref{fig:results_compositions}, we plot \Cref{fig:results_convergences} with a logarithmic scale to capture the huge variations in $t_c$. The shortest simulation corresponds to $\varepsilon_q=0.1625$, $\varepsilon_w=0.9675$, converging at $t_c=65.9$. Conversely, a few simulations along the viability boundary fail to converge before ending at $t=T$; these are plotted in white. Simulations near the viability boundary take the longest to converge, with times $t_c$ decreasing on either side. The shortest times to convergence correspond with the same high-population wedge area seen in \Cref{fig:results_compositions}, reinforcing the connections between these different pieces of the simulation results.

We note that non-viable populations tend to have $t_c$ near the middle of the range. At $\varepsilon_q=\varepsilon_w=1.0$, for example, neither the queen nor the workers lay any eggs, with the initial population slowly dying off at a rate of $\sfrac{1}{\lambda}$ per unit $t$. In such cases, $t_c$ depends mainly on the definition of convergence. Here, denoting the convergence threshold of $10^{-5}$ as $\delta$ and the initial population as $N_0$, we have $t_c=\log_{\frac{\lambda-1}{\lambda}}\frac{\delta\lambda}{N_0}$. When the lifespan $\lambda=\lambda_w=\lambda_d=45$ and $N_0=w_0+d_0=35$, we get $t_c\approx500$ for the non-reproducing populations. It's interesting that higher populations, pressured by the excess deaths for having $N>k$, tend to converge much faster. Lastly, we observe faint `ripples' in the wedge area of the plot, which we dismiss as artifacts of our implementation and definition of convergence.

Taken together, these results serve to illustrate the complex interplay of factors behind arrhenotoky population dynamics. A balancing act is evident, wherein certain combinations of altruism $\varepsilon_w$ and $\varepsilon_q$ yield divergent results in terms of the total populations $N_t$, drone-worker compositions $\sfrac{d}{w}$, and times to convergence $t_c$. We emphasize how the parameter space (or heatmap) is partitioned into distinct regions, including the wedge and other viable/non-viable areas. Next, we use the ODE model equations from \Cref{model} to compute $N_t$ analytically, revealing how these regions are governed by different rules corresponding to three factors which limit the rate of fertilization.

\subsection{Mathematical Analysis}\label{math}

In this section, we calculate equilibrium populations given $\varepsilon_w$ and $\varepsilon_q$; that is, the total population of workers and drones upon convergence $N_{t_{c}}$. This is achieved by solving for the equilibrium $\dot{w}=0$, separately for each of the three branches in the \texttt{min} function of $E_d$ \eqref{eq:dronerep}.

\subsubsection{Limiting Factors}\label{limiting}

Populations can be limited by three distinct factors, which determine the number of eggs $E_d$ that are fertilized by drones during each time step. These factors correspond to the three parameters in the \texttt{min} function of $E_d$ \eqref{eq:dronerep}.
\begin{table}[h]
    \centering
    \caption{Limiting factors}
    \label{tab:limiting}
    \begin{tabular}{c l}
        \toprule
        Symbol & Description \\
        \midrule
        ${J_d}/{u_d}$ & drone fertilization capacity, in terms of energy \\
        $E_q$ & new eggs from the queen, eligible for fertilization \\
        $d$ & drone population, as fertilization results in death \\
        \bottomrule
    \end{tabular}
\end{table}

\textbf{Case 1:} $E_d = \min(E_q,d,\frac{J_d}{u_d}) = \frac{J_d}{u_d}$.

We start by analyzing the case of ${J_d}/{u_d}$, disregarding the difficult case of $d$ for now. Without considering $d$, this case is defined by the condition:
\begin{equation}\label{case1j}
    \frac{J_d}{u_d} \leq \frac{J_q}{u_q}
\end{equation}

Recalling that $J_d = w\varepsilon_w \varepsilon_q$, $J_q = w(1-\varepsilon_q)\varepsilon_w$, and $u_d < u_q$ by assumption, 
\begin{equation}\label{case1ep}
    \varepsilon_q \leq \frac{u_d}{u_q+u_d}
\end{equation}

The right-hand side of \eqref{case1ep} represents the unit cost of reproduction for drones compared to the total unit costs of reproduction for all reproducing individuals. Plugging in the values for $u_d$ and $u_q$ from \Cref{tab:limiting}, we have that
\begin{equation}\label{alpha025}
    \varepsilon_q \leq 0.25
\end{equation}

Of course, we can express the general condition \eqref{case1ep} for any combination of $u_d$ and $u_q$, but for ease of presentation, we will use $\varepsilon_q \leq 0.25$ for the remainder of this section to remain consistent with the simulation results presented in \Cref{fig:results}. 

We find the equilibrium population by solving the worker replicator equation at $\dot{w}=0$ (assuming that $w>0$) for the total population $N=w+d$, following \eqref{eq:workerdeath} for worker deaths:
\begin{eqnarray}
    \dot{w} &=& E_d - D_w = 0 \\
    \frac{J_d}{u_d} &=& \max\left( 1, e^{\frac{N}{k}-1} \right) \frac{w}{\lambda_w}
\end{eqnarray}

For now, we assume that $N \geq k$; that is, $\max\left( 1, e^{\frac{N}{k}-1} \right) = e^{\frac{N}{k}-1}$:
\begin{eqnarray}\label{equilibrium1}
    \frac{w\varepsilon_w\varepsilon_q}{u_d} &=& e^{\frac{N}{k}-1}\frac{w}{\lambda_w} \\
    N &=& k \left( \log\left(\frac{\lambda_w}{u_d}\varepsilon_w\varepsilon_q\right) + 1 \right)
\end{eqnarray}

The equation above applies to region A of the analytical heatmaps in \Cref{fig:analysis}, while also proving that equilibrium populations are proportional to $k$. For the case where $N < k$, see \Cref{viability}.

\textbf{Case 2:} $E_d = \min(E_q,d,\frac{J_d}{u_d}) = E_q = \frac{J_q}{u_q}$.

Next, we solve for the limiting factor $E_q$, defined by the inverse of condition \eqref{alpha025}:
\begin{equation}\label{025alpha}
    \varepsilon_q \geq 0.25
\end{equation}
Note that at the boundary $\varepsilon_q=0.25$, both cases are equivalent. Proceeding through our calculation, the results are similar to \eqref{equilibrium1}, with $J_q$ and $u_q$ replacing $J_d$ and $u_d$:
\begin{equation}\label{equilibrium2}
    N = k \left( \log\left(\frac{\lambda_w}{u_q}(1-\varepsilon_q)\varepsilon_w\right) + 1 \right)
\end{equation}
This equation above applies to region B of the analytical heatmaps in \Cref{fig:analysis}.

\subsubsection{Viability}\label{viability}

The analysis so far has focused on cases where the total equilibrium population exceeds the carrying capacity; that is, $N \geq k$, such that death rates are scaled by a factor of $e^{\frac{N}{k}-1}$. However, some populations fail to reach the carrying capacity $k$. In the simulation results, we observe two distinct states of equilibrium:
\begin{itemize}
    \item Viable: the population reaches $k$, so the exponential death rate takes effect. Eventually, the increasing death rate catches up to the birth rate, resulting in equilibrium. If not for these excess deaths, unbounded population growth would have occurred.
    \item Non-viable: the population never reaches $k$, so the death rate remains at the default of $\frac{1}{\lambda}$. At some point (often $t=0$), the population hits its peak; here, the birth rate is, by definition, not growing nor greater than the death rate. Although we can't rule out the possibility of an equilibrium where $0<N<k$, these populations have trended asymptotically toward zero in all relevant simulations.
\end{itemize}

Therefore, the condition for viability is that the projected equilibrium population is greater than $k$. Based on \eqref{equilibrium1}, we formulate this condition in the case of \eqref{alpha025} where $\varepsilon_q \leq 0.25$:
\begin{equation}
    \log\left(\frac{\lambda_w}{u_d}\varepsilon_w\varepsilon_q\right) > 0
\end{equation}
\begin{equation}
    \varepsilon_w\varepsilon_q > \frac{u_d}{\lambda_w}
\end{equation}
The equation above defines the boundary between regions A and C on the analytical heatmaps in \Cref{fig:analysis}. Similarly, we formulate this condition in the case of \eqref{025alpha} where $\varepsilon_q \geq 0.25$, based on \eqref{equilibrium2}:
\begin{equation}
    \log\left(\frac{\lambda_w}{u_q}(1-\varepsilon_q)\varepsilon_w\right) > 0
\end{equation}
\begin{equation}
    (1-\varepsilon_q)\varepsilon_w > \frac{u_q}{\lambda_w}
\end{equation}
This equation defines the boundary between regions B and D on the analytical heatmaps in \Cref{fig:analysis}; regions C and D consist of non-viable populations, which always converge to zero.

\subsubsection{Boundary Conditions}\label{boundaries}

\textbf{Case 3:} $E_d = \min(E_q,d,\frac{J_d}{u_d}) = d$.

Next, we proceed to the third limiting factor $d$, as defined by the condition:
\begin{equation}\label{conditiond}
    E_d = \min\left(E_q, d, \frac{J_d}{u_d} \right) = d
\end{equation}
We do not have a closed-form solution for the equilibrium population $N$ in this case. Rather than calculating $N$, we proceed by delineating the boundaries of this case in terms of $\varepsilon_w$ and $\varepsilon_q$. This is achieved by formulating the condition \eqref{conditiond} in terms of model parameters, then finding the line along which this inequality is balanced. We perform these calculations twice; first from the direction of case 1, then from the direction of case 2.

Starting with case 1, we reformulate condition \eqref{conditiond} in the case of \eqref{case1j} where $\frac{J_d}{u_d} \leq \frac{J_q}{u_q}$:
\begin{equation}
    d \leq \frac{J_d}{u_d} \leq \frac{J_q}{u_q}
\end{equation}
\begin{equation}\label{dronecondition1}
    \frac{d}{w} \leq \frac{\varepsilon_w\varepsilon_q}{u_d}
\end{equation}

as long as $w>0$. 

The condition is thus defined by the population ratio $\frac{d}{w}$ in relation to the model parameters. This makes sense intuitively; when the number of drones $d$ is the factor limiting population growth, it follows that the ratio of drones to workers must remain low. If the drone population increased, the limiting factor would change to $\frac{J_d}{u_d}$, as in case 1.

\begin{figure}%[H]
    \centering
    
    \begin{subfigure}{.5\textwidth}
        \centering
        \captionsetup{width=.9\linewidth}
        \includegraphics[scale=0.5]{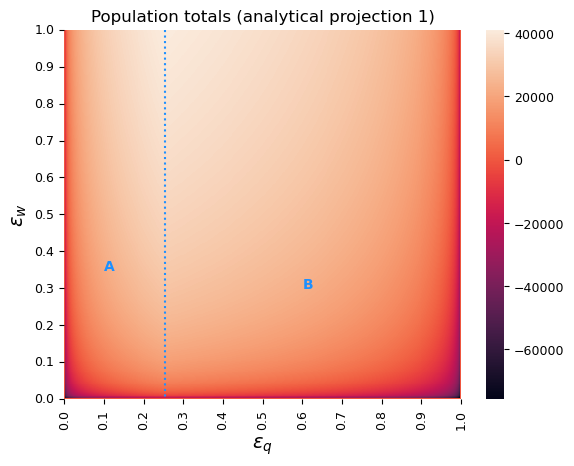}
        \caption{The vertical line at $\varepsilon_q=0.25$ divides the plane between populations limited by drone energy (A) and queen energy (B). We solve for the total population $N$ at equilibrium ($\dot{w}=0$) in regions A and B, assuming $N \geq k$.}
        \label{fig:analysis_ab}
    \end{subfigure}%
    \begin{subfigure}{.5\textwidth}
        \centering
        \captionsetup{width=.9\linewidth}
        \includegraphics[scale=0.5]{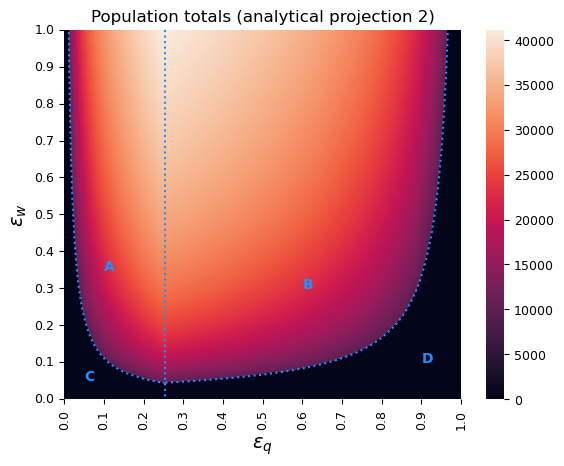}
        \caption{Solving for the viability boundary separates the non-viable regions C and D from the viable regions A and B.\\\\}
        \label{fig:analysis_abcd}
    \end{subfigure}

    \begin{subfigure}{.5\textwidth}
        \centering
        \captionsetup{width=.9\linewidth}
        \includegraphics[scale=0.5]{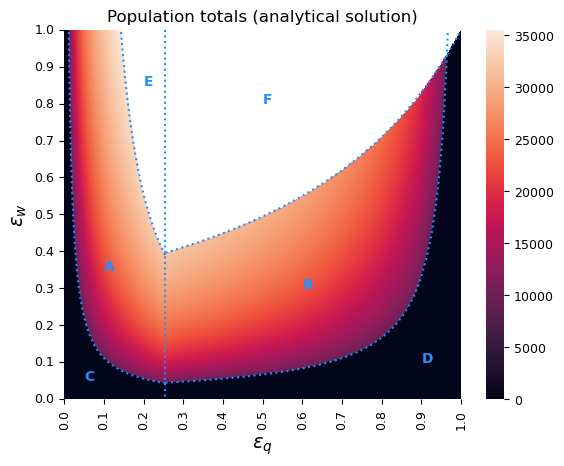}
        \caption{Solving for the limiting factor boundary separates the drone-limited regions E and F from the energy-limited regions A and B. This is the full heatmap of $N$ over the analytically solved regions.}
        \label{fig:analysis_abcdef}
    \end{subfigure}%
    \begin{subfigure}{.5\textwidth}
        \centering
        \captionsetup{width=.9\linewidth}
        \includegraphics[scale=0.5]{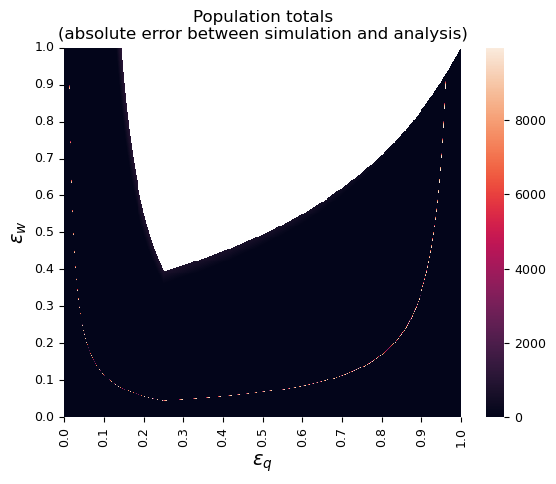}
        \caption{Absolute error between analytical solutions and simulated results over the solved regions. Near-zero everywhere except for the non-converged populations along the viability boundary.}
        \label{fig:analysis_error}
    \end{subfigure}
    
    \caption{Step-by-step illustration: developing an analytical solution for the total equilibrium population $N$, in terms of $\varepsilon_q$, $\varepsilon_w$, and  other model parameters.}
    \label{fig:analysis}
\end{figure}

Although we can't describe the equilibrium population directly, we define this region's boundaries by solving the drone replicator equation at the equilibrium $\dot{d}=0$ for the population ratio $\frac{d}{w}$, assuming $N\leq k$:
\begin{equation}
    \dot{d}=\frac{J_w}{u_w}+\frac{J_q}{u_q}-2\frac{J_d}{u_d}-\frac{d}{\lambda_d}=0
\end{equation}
Note that $u_w = 2 u_q = 6 u_d$ under the simulation conditions presented, as described in \Cref{tab:constants}:
\begin{equation}\label{droneboundary1}
    \frac{d}{w}=\frac{\lambda_d}{u_w}\left(1+\varepsilon_w(1-14\varepsilon_q)\right)
\end{equation}
Finally, we combine \eqref{dronecondition1} and \eqref{droneboundary1} to define the boundary between regions A and E of the analytical heatmaps in \Cref{fig:analysis}:
\begin{equation}   6\varepsilon_q\varepsilon_w=\lambda_d\left(1+\varepsilon_w(1-14\varepsilon_q)\right)
\end{equation}

We now repeat this process from the direction of case 2 in order to define the remaining boundary. To start, we reformulate condition \eqref{conditiond} again, but in the case of \eqref{025alpha} where $\frac{J_q}{u_q} \leq \frac{J_d}{u_d}$:
\begin{equation}
    d \leq \frac{J_q}{u_q} \leq \frac{J_d}{u_d}
\end{equation}
\begin{equation}
    \frac{d}{w} \leq \frac{(1-\varepsilon_q)\varepsilon_w}{u_w}
\end{equation}
as long as $w>0$. 

This condition is analogous to \eqref{dronecondition1} and follows the same intuitive logic. Repeating the above process, we define the boundary between regions B and F of the analytical heatmaps in \Cref{fig:analysis}:
\begin{equation}
    \varepsilon_w(1-\varepsilon_q)=\lambda_d\left(1+\varepsilon_w(2\varepsilon_q-3)\right)
\end{equation}

% spaces to align the nested cases
% https://tex.stackexchange.com/questions/128070/equation-how-to-create-nested-multiple-cases-in-latex-to-align-the-qualifiers

\newlength{\casea}
\setlength{\casea}{\widthof{$k \left( \ln\left(\frac{\lambda_w}{u_q}(1-\varepsilon_q)\varepsilon_w\right) + 1 \right)$} - \widthof{$k \left( \ln\left(\frac{\lambda_w}{u_d}\varepsilon_q\varepsilon_w\right) + 1 \right)$}}

\newlength{\caseb}
\setlength{\caseb}{\widthof{$(1-\varepsilon_q)\varepsilon_w > \frac{u_q}{\lambda_w}$} - \widthof{$\varepsilon_q\varepsilon_w > \frac{u_d}{\lambda_w}$}}

The equilibrium population is thus defined as follows, for all solved cases:
\begin{align}
    N =\begin{cases}
    \begin{cases}
        \begin{cases}
            k \left( \ln\left(\frac{\lambda_w}{u_d}\varepsilon_q\varepsilon_w\right) + 1 \right) & \hspace{\casea} \text{if $\varepsilon_q\varepsilon_w > \frac{u_d}{\lambda_w}$} \\
            0 & \hspace{\casea} \text{otherwise} \\
        \end{cases} & \hspace{\caseb} \text{if $6\varepsilon_q\varepsilon_w < \lambda_d(1+\varepsilon_w(1-14\varepsilon_q))$} \\
        \text{unsolved} & \hspace{\caseb} \text{otherwise}
    \end{cases} & \text{if $\varepsilon_q \leq 0.25$} \\
    \begin{cases}
        \begin{cases}
            k \left( \ln\left(\frac{\lambda_w}{u_q}(1-\varepsilon_q)\varepsilon_w\right) + 1 \right) & \text{if $(1-\varepsilon_q)\varepsilon_w > \frac{u_q}{\lambda_w}$} \\
            0 & \text{otherwise} \\
        \end{cases} & \text{if $\varepsilon_w(1-\varepsilon_q) < \lambda_d(1+\varepsilon_w(2\varepsilon_q-3))$} \\
        \text{unsolved} & \text{otherwise}
    \end{cases} & \text{otherwise} \\
    \end{cases}
\end{align}

\section{Discussion}

Our results serve to illustrate the impact of altruism on arrhenotokous populations. A delicate balancing act is evident, where populations may be limited by various factors depending on the patterns of altruism at play.

Altruism, at some level, is necessary for survival in all populations. Given that workers collect energy on behalf on the entire hive, altruism enables their reproduction via the queen (recall that only the queen's eggs can be fertilized, becoming workers). Otherwise, workers lay eggs that only hatch as drones, preventing the hive from gathering more energy as workers die off without replacement. Hence, all populations where $\varepsilon_w=0$ are non-viable.

Queens must also act with some degree of altruism in order for the populations to survive. If the queen hoards all energy provided by her workers, drones are left without enough energy to fertilize her eggs, jeopardizing the worker population in a similar manner. Conversely, if the queen acts with total selflessness, she lacks the energy required to lay any eggs of her own. Therefore she performs a balancing act, such that populations where $\varepsilon_q=0$ or $\varepsilon_q=1$ are both non-viable.

Trade-offs are introduced into the model by the nature of drone reproduction. Recall that two pathways exist for the birth of drones: unfertilized eggs laid by the queen, and eggs laid by workers (which cannot be fertilized). This dynamic gives the workers room to act less altruistically, with minimal effect on the total population, as long as the queen acts altruistically enough to enable the drones to fertilize. This explains the high-population ridge across the heatmap in \Cref{fig:results_totals}, curving from $\varepsilon_q=0.15$ to $\varepsilon_q=0.25$.

If, however, the workers and queen both act with excessive altruism, energy is wasted as the drones are allocated more than they're able to use. Given that fertilization results in death, there are simply not enough drones in existence to spend their fertilization `budget'. This excess energy could have been used by either the workers or the queen to lay more eggs and increase the drone population; therefore, such populations cannot be optimal. It is precisely this region of the heatmap that was not described analytically, given that $\dot{w}$ \eqref{wdot} depends on $\dot{d}$ \eqref{ddot} in these cases.

\subsection{Evolution of Altruism}

We emphasize that our model frames altruism in arrhenotoky as a resource allocation problem. Worker bees collect fixed amounts of energy per time step, which is allocated per $\varepsilon_w$ and $\varepsilon_q$, then consumed during reproduction or else wasted. What our model truly measures is the energy efficiency of hives, across the two-dimensional parameter space of queen and worker altruism. Thus, the notion of `population' is merely a convenient abstraction through which our results are framed. Higher-efficiency beehive configurations can support greater multiples of the imposed carrying capacity $k$.

Within this framework, it makes sense to consider the parameters giving rise to the highest populations as `optimal' beehive configurations. Whereas high populations are not necessarily optimal under real-world conditions, our model directly equates population/efficiency with evolutionary fitness. The maxim `survival of the fittest' dictates that real-world evolution converges toward the state of greatest fitness. With that in mind, our simulation's optimum (at $\varepsilon_w=1.0$ and $\varepsilon_q=0.13$) sheds light on \textit{why} altruism evolved and \textit{to what extent}, if not \textit{how} the process played out in particular. Our model considers reproduction, but not the degree of genetic relatedness among individuals. Therefore, our results neither claim nor disclaim kin selection theory, but provide a mathematical explanation for the evolution of altruism in certain contexts.

\subsection{Future Directions}

We offer several extensions to this work, involving new additions to the model or the relaxation of assumptions. For instance, beehive populations are profoundly affected by seasonal fluctuations, through changes in the abundance of food, as well as the role that drones play in hive temperature regulation. The dynamic nature of queen bees is another unaddressed question; while one queen suffices for short simulations, the queen-replacing processes of supersedure and swarming are essential for describing longer-term beehive development and persistence \cite{rowland_1987}. Our model also assumes fixed rates of altruism within each hive, but the introduction of behavior-altering mutations may bring about collective action problems and other interesting game-theoretic dynamics. Lastly, given that our model can already describe other arrhenotokous species, we might consider further generalizations to other eusocial species, in pursuit of a more comprehensive understanding of altruistic behavior.

\section{Conclusion}

We developed a population dynamical model of arrhenotoky, simulating beehive population dynamics across all levels of altruism in the queen and workers. Our results show that levels of altruistic behaviors are among the primary determinants of beehive success. The general correlation between worker altruism $\varepsilon_w$ and population size $N_T$ serves to demonstrate the remarkable evolutionary advantages conferred by eusociality. A hallmark of eusociality is the distinction between reproductive and non-reproductive castes, which is represented in our model by the total lack of eggs laid by workers in simulations where $\varepsilon_w=1.0$. Furthermore, we show why modest altruism from the queen is also necessary for the hive's survival, with successful populations operating under a delicate balancing act. Taken together, our results provide a mathematical explanation for the real-world co-evolution of altruism, arrhenotoky, and eusociality.

\section{Acknowledgments}

The authors thank Daniel Cooney and Feng Fu for helpful comments and discussions. O.J.C. was supported by the Neukom Institute for Computational Science at Dartmouth College when this work was done.

\FloatBarrier
\newpage

\bibliographystyle{unsrt}
\bibliography{references}

\newpage

\section{Supplementary Information}

\subsection{Simulation Details}

The beehive population model is implemented in python as a system of ordinary differential equations, using the \verb|solve_ivp| method from \verb|scipy| to solve the replicator equations $\dot{w}$ \eqref{wdot} and $\dot{d}$ \eqref{ddot} until time $T$. A grid search is conducted across 160,681 ($401^2$) configurations of $\varepsilon_w$ and $\varepsilon_q$, recording three results from each instance: the total population, the ratio of drones to workers, and the time of population convergence. Computation is sped up greatly by multiprocessing. To display these results, population heatmaps over the space of $\varepsilon_w$ and $\varepsilon_q$ are generated using \verb|seaborn|. We publish our implementation on GitHub: \url{https://github.com/ZackNathan/arrhenotoky_simulation}.

\subsection{Supplemental Figures}\label{supplement}

Here, we include additional heatmaps of the simulation results, comparable to those in \Cref{fig:results}. We vary the model parameters from \Cref{tab:constants} to illustrate their effects; in particular, we focus on the unit costs of reproduction $u_w$, $u_q$, and $u_d$. Notably, changing these values shifts the boundaries between analytical regions (see \Cref{fig:analysis_abcdef}), without altering the general pattern.

\vspace{1em}

\begin{figure}[h!]
    \centering
    
    \begin{subfigure}{.5\textwidth}
        \centering
        \captionsetup{width=.9\linewidth}
        \includegraphics[scale=0.5]{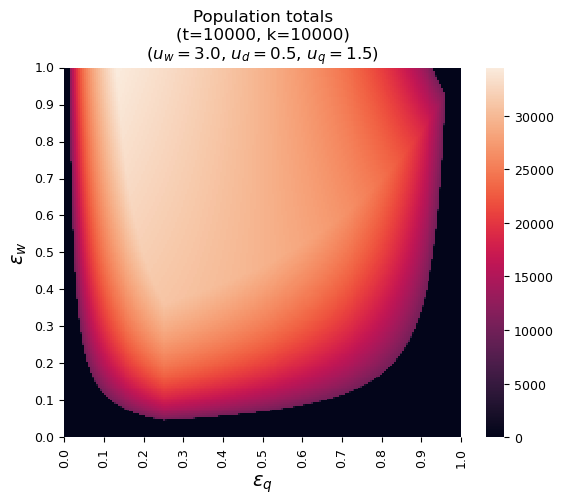}
    \end{subfigure}%
    \begin{subfigure}{.5\textwidth}
        \centering
        \captionsetup{width=.9\linewidth}
        \includegraphics[scale=0.5]{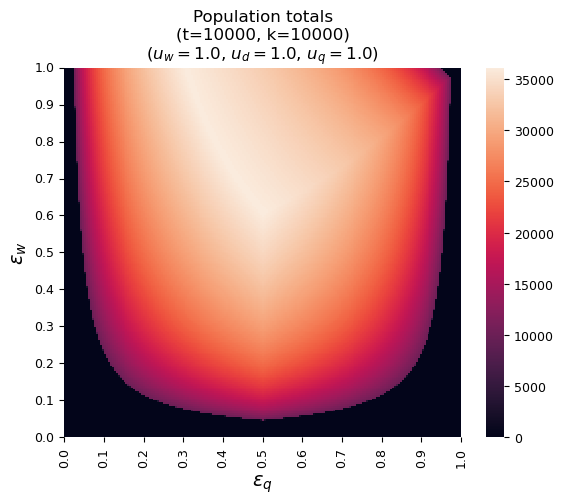}
    \end{subfigure}

    \begin{subfigure}{.5\textwidth}
        \centering
        \captionsetup{width=.9\linewidth}
        \includegraphics[scale=0.5]{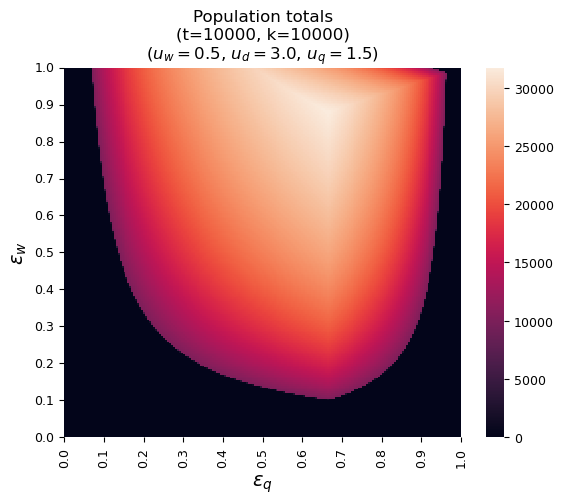}
    \end{subfigure}%
    \begin{subfigure}{.5\textwidth}
        \centering
        \captionsetup{width=.9\linewidth}
        \includegraphics[scale=0.5]{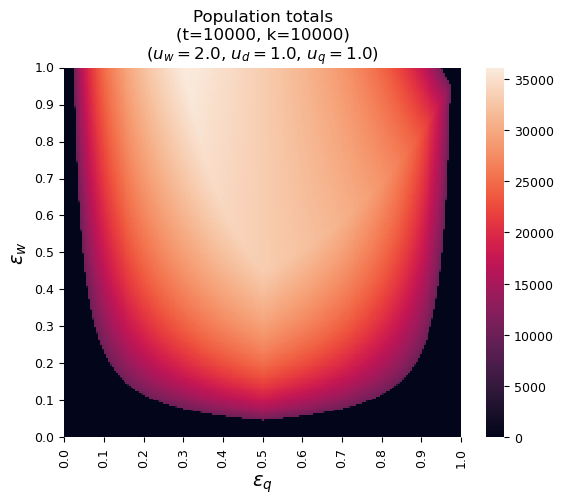}
    \end{subfigure}
    
    \caption{Heatmaps showing the final population totals $N_T$ with varying unit costs of reproduction $u_w$, $u_q$, and $u_d$. The upper left plot shows results with the default parameter values, equivalent to \Cref{fig:results_totals}. We denote the parameter values in the titles of each plot.}
    \label{fig:supplement_1}
\end{figure}

\begin{figure}
    \centering
    
    \begin{subfigure}{.5\textwidth}
        \centering
        \captionsetup{width=.9\linewidth}
        \includegraphics[scale=0.5]{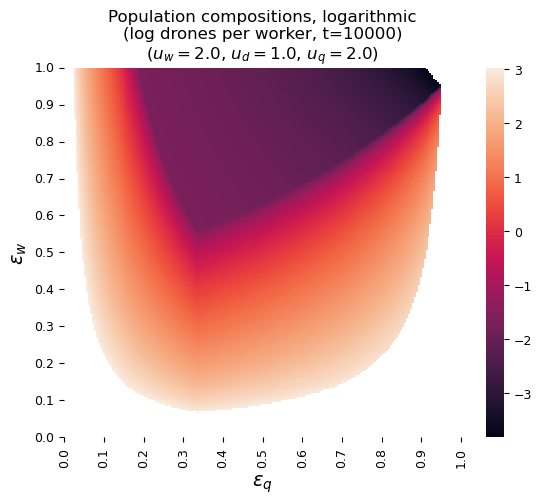}
    \end{subfigure}%
    \begin{subfigure}{.5\textwidth}
        \centering
        \captionsetup{width=.9\linewidth}
        \includegraphics[scale=0.5]{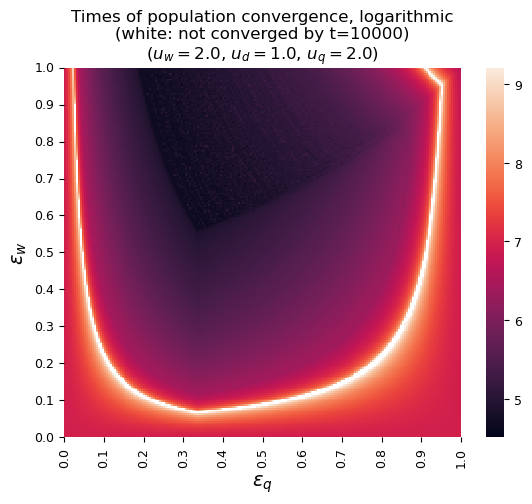}
    \end{subfigure}

    \begin{subfigure}{.5\textwidth}
        \centering
        \captionsetup{width=.9\linewidth}
        \includegraphics[scale=0.5]{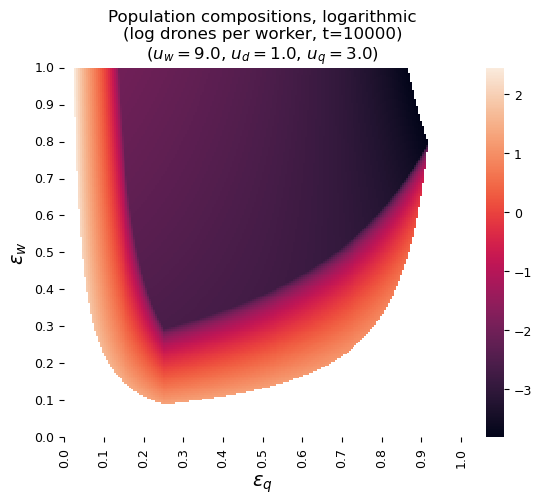}
    \end{subfigure}%
    \begin{subfigure}{.5\textwidth}
        \centering
        \captionsetup{width=.9\linewidth}
        \includegraphics[scale=0.5]{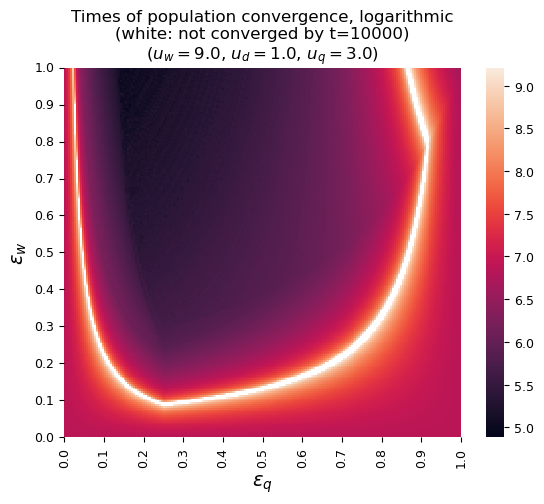}
    \end{subfigure}

    \begin{subfigure}{.5\textwidth}
        \centering
        \captionsetup{width=.9\linewidth}
        \includegraphics[scale=0.5]{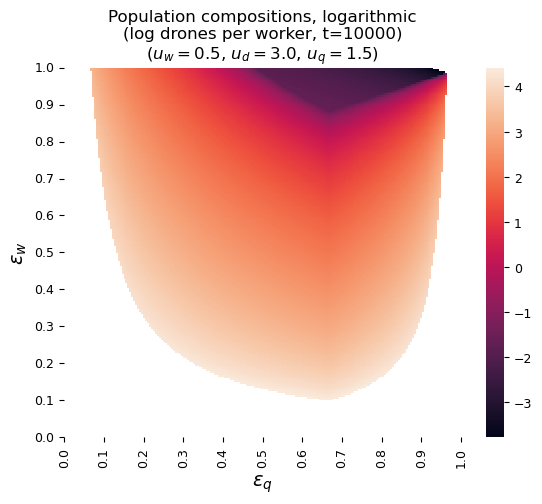}
    \end{subfigure}%
    \begin{subfigure}{.5\textwidth}
        \centering
        \captionsetup{width=.9\linewidth}
        \includegraphics[scale=0.5]{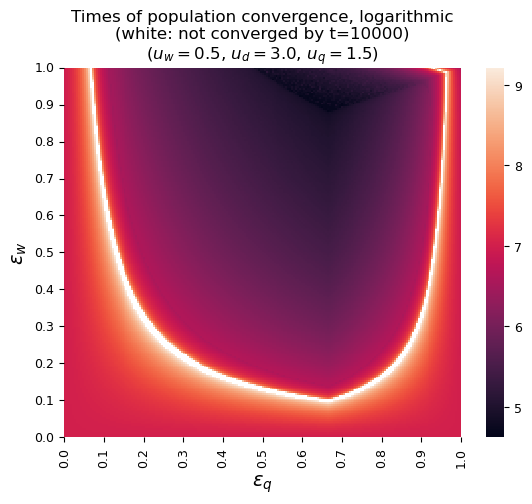}
    \end{subfigure}
    
    \caption{Heatmaps showing the population compositions $\sfrac{d}{w}$ and times of convergence $t_c$ with varying unit costs of reproduction $u_w$, $u_q$, and $u_d$. These results serve to illustrate how divergent parameter values yield heatmaps with diverse shapes and ranges, but always maintaining the same overall pattern of regions seen in \Cref{fig:analysis_abcdef}. We denote the parameter values in the titles of each plot.}
    \label{fig:supplement_2}
\end{figure}

\begin{figure}
    \centering
    
    \begin{subfigure}{.5\textwidth}
        \centering
        \captionsetup{width=.9\linewidth}
        \includegraphics[scale=0.5]{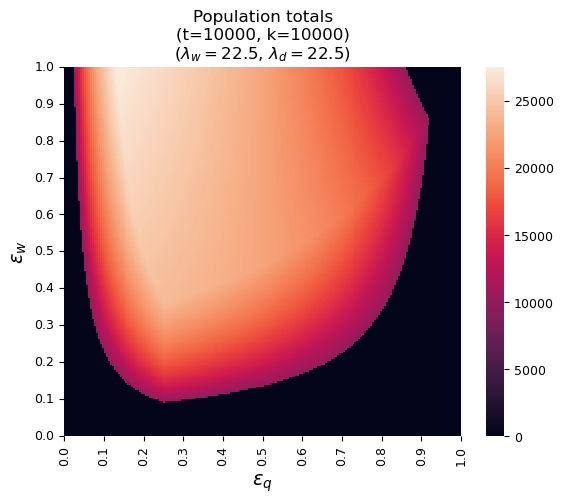}
    \end{subfigure}%
    \begin{subfigure}{.5\textwidth}
        \centering
        \captionsetup{width=.9\linewidth}
        \includegraphics[scale=0.5]{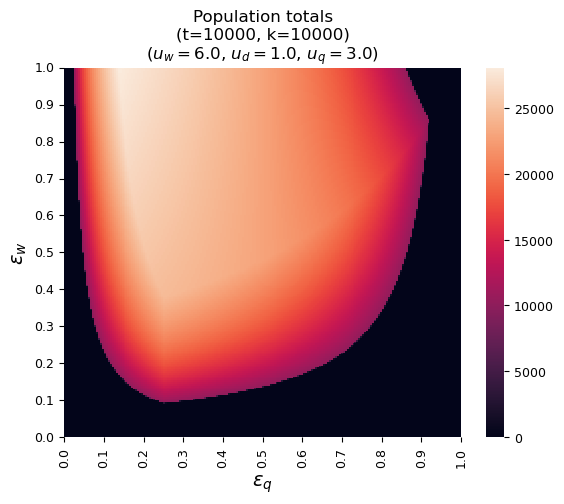}
    \end{subfigure}

    \begin{subfigure}{.5\textwidth}
        \centering
        \captionsetup{width=.9\linewidth}
        \includegraphics[scale=0.5]{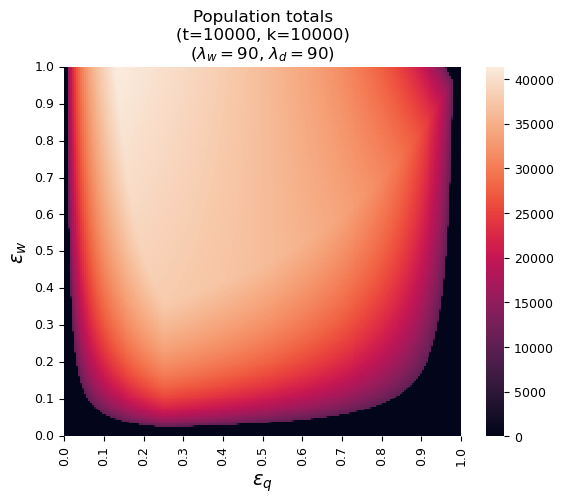}
    \end{subfigure}%
    \begin{subfigure}{.5\textwidth}
        \centering
        \captionsetup{width=.9\linewidth}
        \includegraphics[scale=0.5]{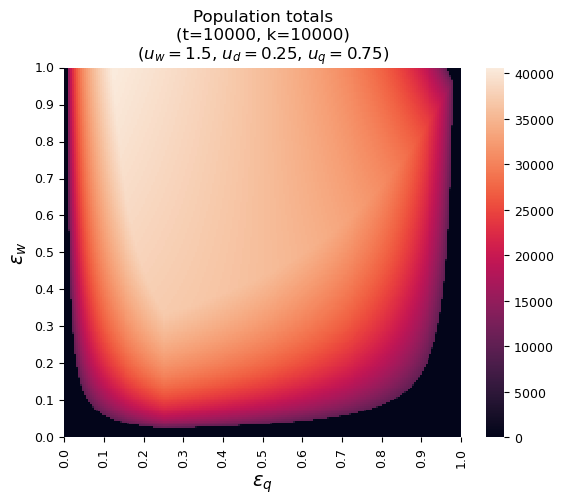}
    \end{subfigure}

    \begin{subfigure}{.5\textwidth}
        \centering
        \captionsetup{width=.9\linewidth}
        \includegraphics[scale=0.5]{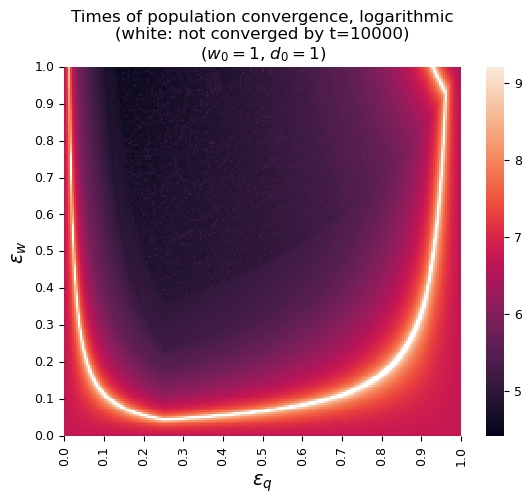}
    \end{subfigure}%
    \begin{subfigure}{.5\textwidth}
        \centering
        \captionsetup{width=.9\linewidth}
        \includegraphics[scale=0.5]{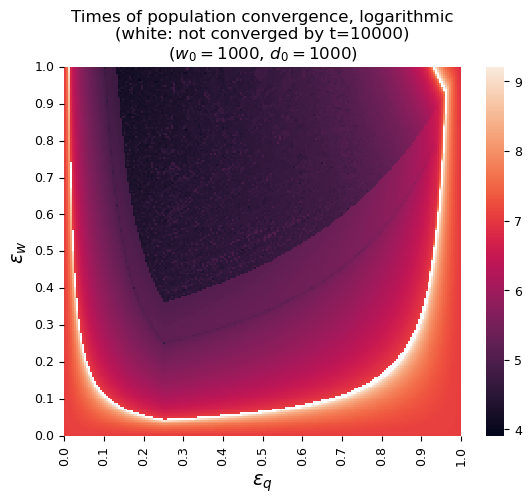}
    \end{subfigure}
    
    \caption{We show that halving the lifespans ($\lambda_w$ and $\lambda_d$) is equivalent to doubling the unit costs; that doubling the lifespans is equivalent to halving the unit costs; and that changing the initial populations ($w_0$ and $d_0$) affects only the times of convergence, and only slightly at that.}
    \label{fig:supplement_3}
\end{figure}

\end{document}